\documentclass[journal]{IEEEtran}

\usepackage{amsmath}
\usepackage{amssymb}
\usepackage{amsfonts}
\usepackage{url}
\usepackage{hyperref}
\usepackage[plain]{fancyref}
\usepackage{graphicx}
\usepackage{diagbox}
\usepackage{mathrsfs}
\usepackage[version=3]{mhchem}
\usepackage{varwidth}
\usepackage{tikz}
\usepackage{multirow}
\usepackage{tabularx}
\usepackage{booktabs}
\usepackage{siunitx}
\usepackage{wasysym}
\usepackage{nicefrac}
\usepackage{color, soul}

\graphicspath{{./fig/}}
\begin{document}

\title{Low Activity Tritium Detection in CCDs Using Deep Learning Techniques}

\author{E.~Rofors, R.~Heller, R.J.~Cooper, J.~Estrada,  G.~Moroni, B.~Nachman, K.~Spears  \\[-5.0ex]
\IEEEaftertitletext{\vspace{-5\baselineskip}}

\thanks{Manuscript received \today. This work was performed under the auspices of the U.S. Department of Energy by Lawrence Berkeley National Laboratory (LBNL) under Contract DE-AC02-05CH11231. The project was funded by the U.S. Department of Energy, National Nuclear Security Administration, Office of Defense Nuclear Nonproliferation Research and Development.}
\thanks{E.~Rofors, R.~Heller, R.J.~Cooper, B.~Nachman, and K.~Spears are at the Lawrence Berkeley National Laboratory, Berkeley, CA 94720, USA. G.~Moroni and J.~Estrada are at the Fermi National Accelerator Laboratory, PO Box 500, Batavia IL 60510, USA}%
}

\maketitle
\pagestyle{empty}
\thispagestyle{empty}

\begin{abstract}
This study explores the use of charge-coupled devices (CCDs) for detecting low-energy beta particles from tritium decay—a critical signal for nuclear safety, nuclear nonproliferation, and environmental monitoring. We employ a dual approach utilizing both measured CCD data and detailed Geant4 simulations. Our analysis compares classical techniques with advanced deep learning methods, including convolutional neural networks (CNNs), autoencoders trained exclusively on tritium data, and preliminary studies on boosted decision trees (BDTs). The CNN, trained on mixed signal/background datasets, demonstrates superior classification performance, while the autoencoder shows the potential of unsupervised, background-agnostic strategies when background characteristics are poorly defined. These results highlight the excellent sensitivity achievable thanks to the background rejection made possible by information-rich CCD data, paving the way for improved portable tritium monitoring.

\end{abstract}

\section{Introduction} \label{sec:intro}

\IEEEPARstart{T}{he} detection and quantitative analysis of tritium ($^3$H) produced in nuclear fission is important for mission areas including safeguards, non-proliferation, emergency response, and consequence management. Tritium, with a half-life of about 12.3 years, decays via beta emission producing a low-energy electron, up to 18.6\,keV, and an antineutrino. Detecting this low-energy beta particle poses a considerable challenge with conventional detectors, as its range in most materials is limited to just a few microns. 
For a goal of in situ measurements, discerning small specific activities among high environmental backgrounds becomes an additional  major challenge.
Tritium detection and quantification is thus typically achieved only with laborious laboratory techniques not suitable for detection in the field~\cite{hou_critical_2008, mao_measurement_2024}.

The most commonly used modality for tritium detection are liquid scintillator counters (LSC), which rely on cocktails of aqueous samples mixed directly with the liquid scintillator~\cite{hofstetter_field_1999-1,horrocks_applications_2012}. This mixing has the advantage facilitating direct interactions of the beta rays with the active material, which would otherwise limited by the extremely short beta ray range. LSC approaches have achieved good sensitivity of order 1 Bq/L, or 0.1 Bq/L with additional electrolytic enhancement~\cite{plastino_tritium_2007,lin_electrolytic_2020,furuta_measurement_2014}. However, the single-use cocktails rely on consumable liquid scintillator, generating significant radioactive waste over time, and requiring more complex logistical support in a deployed remote system. Backgrounds from other low energy beta emitters like ($^{14}$C) are also difficult or impossible to distinguish from tritium~\cite{mao_measurement_2024}.

Other approaches include gas proportional counters, which have achieved impressive sensitivity, but are bulky systems relying on significant passive and active shielding, and require operation at high voltage (3000\,V)~\cite{bowman_proportional_1981,stanga_improved_2006}. Solid state detectors including silicon avalanche detectors~\cite{street_ciency_nodate} and plastic scintillators are also useful for moderate tritium concentrations~\cite{uda_detection_2010, sanada_development_2024}, but typically have very low detection efficiency due to absorption in the inactive outer layers of the detector.
For quantification of low tritium activities ($<1$ pCi/m$^3$) in the field, a highly efficient detector that can operate with minimal shielding, infrastructure, and logistical intervention is needed.

Towards the goal of in-situ tritium detection in fieldable systems, we present the use of thick, fully depleted scientific CCDs~\cite{holland2003}. These sensors provide precise energy measurements, few-electron noise, and fine pixelation on the order of 10 to 15~microns. Crucial for the tritium application, back-illuminated CCDs with ultra thin entrance windows can efficiently collect all charge deposited deeper than the first 10 to 100\,nm of the sensor volume~\cite{holland_fully_2023,goldschmidt_veryfastccd_2023}. CCD exposures result in detailed, information-rich tracks for each interacting particle. Based on the energy deposition and track or cluster shapes, these interactions can be associated with tritium beta ray or other background processes. As background interaction are generally much more common than tritium beta rays, classification with strong background rejection is especially important. Deep learning offers promising tools to obtain excellent signal and background separation and extend sensitivity to low tritium activities even with modest counting times.

To develop this approach, we utilize data collected with a CCD detector and leverage recent advancements in machine learning. Previous work has demonstrated the potential of deep learning for particle track classification in CCDs\cite{britt2022}. Building on these advances, we explore both supervised and unsupervised approaches for enhanced tritium detection.

The envisioned deployment of this detection modality was inspired by the IARPA GRAIL program~\cite{grail} for autonomous, in situ tritium monitoring, targeting daily quantification of atmospheric tritium levels. The GRAIL workflow would involve collection and isolation of hydrogen gas (including tritium) from atmospheric samples, conversion to water, and deposition of this water directly on or near the CCD surface for quantification over 24~hours. In this paper, we focus on the quantification step and seek to understand the tritium and background separation power achievable with CCD detectors and machine learning analysis. The GRAIL quantification goals include minimum detectable activities of 0.1\,pCi of tritium per standard cubic meter (SCM) of air with a 24 hour count time, and a stretch goal of 0.05\,pCi/SCM sensitivity in 12 hours.

In the following sections, we describe our experimental setup, simulation framework, and the advanced analysis methods—including both classical and deep learning techniques employed to enhance tritium signal identification.

\section{Measurement \label{sec:measurement}}

Two distinct CCD datasets were collected for our study using a CCD operated at 140\,K with a substrate voltage of 70\,V. The sensor has an active area of 96$\times$16.5\,mm$^2$ with a thickness of 625\,\si{\micro\meter} and features high spatial resolution with $\sim15\cdot15$\,\si{\micro\meter}$^2$ pixels. The sensor was fabricated with a minimal inactive surface layer of roughly 10\,nm, corresponding to a quantum efficiency for tritium beta rays of about 60\%. The system was operated in vacuum.

The first dataset consists of background measurements collected over approximately 25\,hours, yielding roughly 6\,000~particle tracks. The measurement was divided over multiple exposures, each lasting approximately 18~minutes.

The second dataset was collected over 37\,hours with a tritium source positioned 3$''$ from the detector face, resulting in 1.6M~tracks. For these measurements, each exposure lasted approximately 3~minutes and 39~seconds.

\begin{figure}
    \centering
    \includegraphics[width=0.8\linewidth]{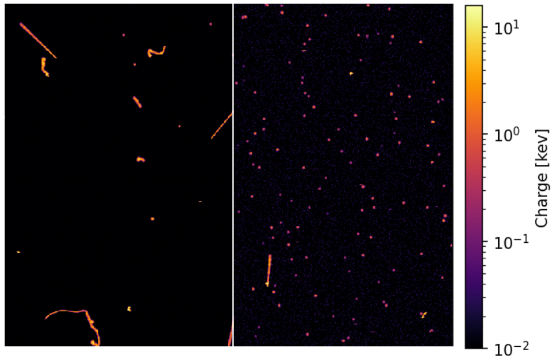}
    \caption{Particle tracks measured by a CCD without (left) and with (right) a tritium source placed 3$''$ from the detector.}
    \label{fig:tracks}
\end{figure}

\begin{figure}
    \centering
    \includegraphics[width=0.8\linewidth]{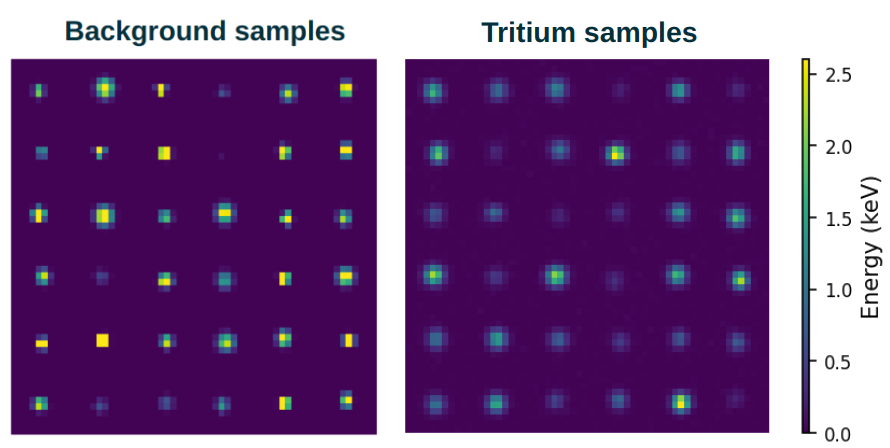}
    \caption{Zoomed in 10$\times$10~pixel views of particle clusters from background (left) and tritium (right).}
    \label{fig:comparison}
    
\end{figure}

Fig.~\ref{fig:tracks} presents 300$\times$500~pixel (4.5$\times$7.5\,mm$^2$) slices of exposure images: the left image shows background radiation over an 18-minute period, and the right displays an exposure in the presence of the tritium source. Some muon tracks are visible in both exposures along with smaller energy deposition clusters. Fig.~\ref{fig:comparison} shows zoomed in views of tritium and background clusters centered around the highest intensity pixel of each cluster. 

To identify and analyze each particle event, we apply a clustering algorithm to identify connected pixels above a threshold of 51\,eV, corresponding to 4$\sigma$ of the noise observed in our measured data. This threshold is large enough to eliminate essentially all false hits arising from noise, but is still small compared to the typical tritium beta ray deposition in keV. For each cluster, we extract a 10$\times$10~pixel region centered around the highest intensity pixel. While some background particle tracks extend beyond this window, tritium clusters are well-contained within this size. Some examples of extracted tritium and background clusters are shown in Fig.~\ref{fig:comparison}. The range of tritium electrons in the detector is much smaller than the size of the CCD pixels but due to drift and diffusion effects of the freed charges, the resulting clusters from tritium electrons typically spread across approximately 20~pixels. Clusters larger than 10$\times$10~pixels are trivial to distinguish from tritium signals, making this window size sufficient for our classification task.

\section{Simulations \label{sec:simulations}}

To complement the measured data and gain deeper insight into the CCD response, we performed extensive Geant4~10.5~\cite{geant4_1} simulations of particles interacting with the CCD. The CCD was modeled as a 600\,\si{\micro\meter} thick block of silicon in vacuum. We used the Livermore low-energy physics list along with decay and radioactive decay physics modules, with atomic deexcitation processes (fluorescence, Auger) enabled. To accurately track low-energy particles near the detector surface, a maximum step size of 20\,nm was enforced in the first 20\,\si{\micro\meter} of the silicon volume.

\subsection{Electron Range}
\begin{figure}[ht]
    \centering
    \includegraphics[width=0.70\linewidth]{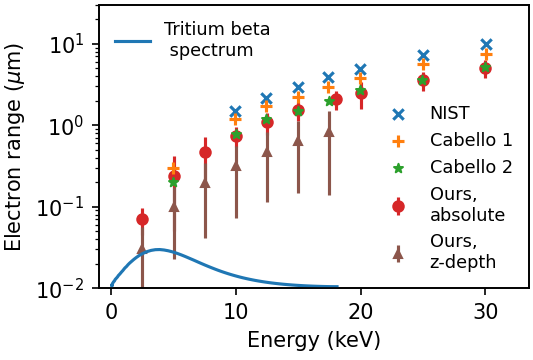}
    \caption{Electron ranges in silicon from NIST~\cite{NIST}, simulations by Cabello et. al.~\cite{cabello2010}, and our Geant4 simulation. The red points show the full path length traveled, while the brown points show the maximum depth reached in the sensor for isotropic emission. The energy spectrum of tritium is overlaid for reference.}
    \label{fig:ranges_data}
\end{figure}

To study the range of electrons in the CCD we fired parallel beams of electrons perpendicular to the surface of the CCD. Fig.~\ref{fig:ranges_data} shows a comparison of electron ranges in silicon between NIST\cite{NIST}, Cabello et al.\cite{cabello2010}, and our simulation. The range of a particle is defined as the straight line distance from the starting point to final absorption point. 

The discrepancies between methods arise from differences in the physical processes included. The NIST data, extending only down to 10\,keV, considers continuous slowing-down approximation based on stopping power alone, without accounting for stochastic energy loss fluctuations or multiple scattering effects. The Cabello~1 curve replicates these NIST conditions in Geant4 by disabling scattering and energy straggling, yielding close agreement with NIST where they overlap. Both our simulation and Cabello~2 include the full physics of multiple Coulomb scattering and fluctuating energy losses (straggling).

\begin{figure}[ht]
    \centering
    \includegraphics[width=0.95\linewidth]{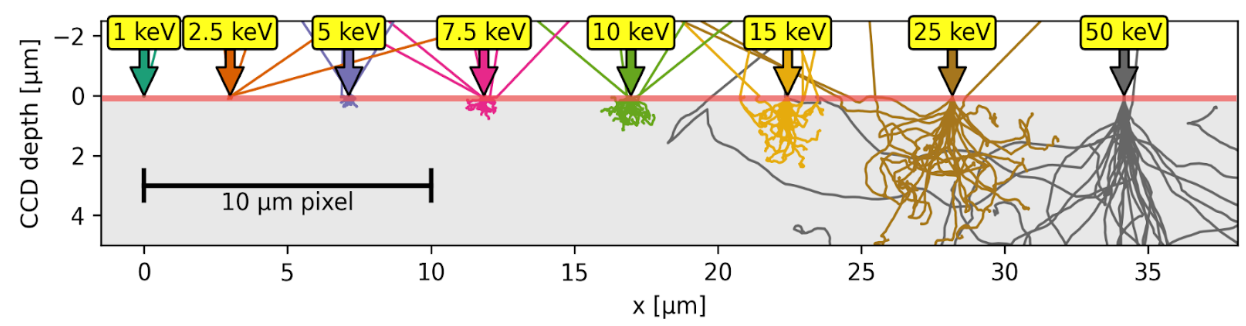}
    \caption{Simulated electron trajectories in a CCD at different incident energies (1--50 keV). The gray region represents the detector volume, with a 10\,\si{\micro\meter} pixel width shown for scale. The red band illustrates a 10\,nm thick dead layer of insensitive detector.}
    \label{fig:ranges_illustrative}
\end{figure}

For CCD tritium detection, the critical parameter is not the particle track length but rather the maximum penetration depth into the detector. If an electron deposits most of its energy in the outermost dead layer (where ohmic polysilicon contacts are located\cite{holland2003}), the event will not be registered. The average maximum penetration depth for isotropically incident electrons on the CCD surface is shown as brown triangles and spans $\sim$0--1\,\si{\micro\meter} for the energy range of tritium beta particles. These depths are significantly shorter than the total path lengths due to the curved, tortuous trajectories. The error bars indicate $\pm$ one standard deviation.
Fig.~\ref{fig:ranges_illustrative} illustrates simulated electron trajectories with incident energies ranging from 1 to 50\,keV. Lower energy electrons (1--7.5\,keV) are characterized by shallow penetration depths, while those with energies between 10 and 50\,keV penetrate further and experience increased scattering.  These simulations help explain the energy-dependent spatial diffusion observed in the experimental data.

\begin{figure}
    \centering
    \includegraphics[width=0.8\linewidth]{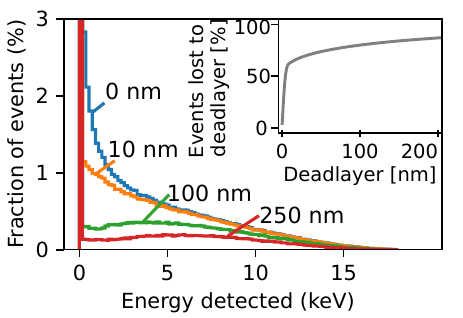}
    \caption{The fraction of tritium beta particles that deposit energy in the sensitive CCD volume for dead layer thicknesses of 0, 10, 100, and 250\,nm. The inset plot show the percentage of incoming tritium particles that are lost due to the dead layer.}
    \label{fig:deadlayers}
\end{figure}

In the manufacturing of CCDs, a thin layer of polysilicon is used as an ohmic contact on the surface that will be facing the tritium source. Because of the weak penetrating power of tritium beta particles, the thickness of this insensitive dead layer must be made as thin as possible. Fig.~\ref{fig:deadlayers} shows the fraction of energy deposited in the dead layer for a tritium source that emits isotropically into the hemisphere on the CCD surface. With a 100\,nm thin dead layer, $\sim$80\% of tritium beta particles are lost, leaving only a fraction as candidates for detection. These type of contacts can be manufactured down to $\sim$10\,nm thin\cite{holland2003}. Although monocrystalline silicon and polysilicon may exhibit slightly different densities, the simulations presented here consider both materials as monocrystalline silicon for simplicity in the analysis. This assumption should not impact the overall conclusions drawn from these results.

\subsection{CCD Charge Drift and Diffusion Modeling \label{sec:diffusion}}
Drift and diffusion affects the charges in a CCD as they move towards the readout plane. These effects were modeled in Python using the energy deposition for each particle step taken in the Geant4 simulation.
The charges liberated by the incoming particles were sampled from a Gaussian distribution with a mean of the deposited $\text{energy} / 3.6\text{\,eV}$ (the mean energy for creating an electron-hole pair) and a variance corresponding to a Fano Factor of 0.16\cite{Janesick1987}. The cloud of charges will drift towards the readout side of the CCD based on the applied voltage. We matched our experimental setup with 70\,V applied to the CCD, resulting in drift times between 0-40\,ns from energy deposition to charge collection, depending on the depth of interaction in the CCD. The drift time is used to apply lateral diffusion that occurs in CCDs. A table of pre-calculated diffusion kernels for varying depth of energy deposition was prepared and applied to the charges, with an increasingly wide Gaussian point spread function with increasing interaction depth. The widths are calculated from the drift times which are fitted from measured responses from CCDs manufactured at Lawrence Berkeley National Laboratory. A per-pixel noise was added to the final image to account for the noise in the CCD. The noise was sampled from a Gaussian distribution with a mean of 0 and a standard deviation of 3.5~electrons as observed in our experimental data.

\subsection{Ice-layer Simulations \label{sec:ice-layer}}
Because of the short range of tritium beta particles, samples should be placed close to the CCD to maximize the signal. One option is to freeze an aqueous sample onto the CCD surface. Here we use ice-layer simulations to investigate how varying thicknesses affect the fraction of tritium beta particles that reach the CCD.

We conducted simulations of tritium-containing ice layers of varying thicknesses placed directly on the CCD surface. 
As beta particles are emitted isotropically within the ice, only a fraction reach the detector with sufficient energy to register a signal.

\begin{figure}
    \centering
    \includegraphics[width=0.85\linewidth]{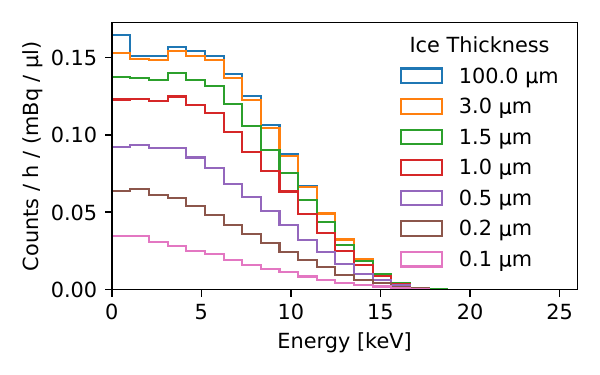}
    \caption{Energy deposited from freezing different thicknesses of ice (0.1-100\,\si{\micro\meter}) from a water sample containing tritium onto a CCD. The units are counts per hour per mBq/µl of tritium in the water sample per 1\,keV wide bins. As ice thickness increases beyond approximately 3\,\si{\micro\meter}, the detection rate shows diminishing returns due to self-absorption of the beta particles within the ice. Only counts that deposit energy in the CCD are shown.}
    \label{fig:ice-layer}
\end{figure}

Fig.~\ref{fig:ice-layer} shows the energy deposition spectrum in the CCD for different ice thicknesses. For very thin layers (0.1-0.2\,\si{\micro\meter}), the number of detected events scales approximately linearly with thickness. However, as the ice layer exceeds a few micrometers, we observe diminishing returns in detection efficiency. This occurs because the range of low-energy beta particles from tritium decay is limited in ice, similar to the penetration depth data (brown triangles) in Fig.~\ref{fig:ranges_data} for silicon. Beta particles emitted deeper in the ice are stopped within the ice itself before reaching the CCD surface. The 100\,\si{\micro\meter} ice layer shows nearly identical performance to the 3\,\si{\micro\meter} layer, indicating that additional thickness beyond this point primarily contributes self-absorption rather than increased detection probability. For an adequately thick ice sample, the expected rate of tritium events is simply determined by the tritium concentration in the original sample.
For layers of ice of at least 3\,\si{\micro\meter} thickness over the area of the CCD, the total count rate will be about 1.57~counts per hour per mBq/µl of tritium in the water sample. We call this factor the tritium deposition factor, $f_{t}$. This factor is a convenient way to express the probability for a tritium beta ray to deposit charge in the sensitive volume of the CCD. For an ice layer of exactly 3~microns thickness, the average probability is about 1.5\%, though it varies significantly throughout the ice.

Depending on the tritium concentration that can be achieved during sample preparation and the surface area of the CCD, these results give an estimate of the expected tritium count rate that can be achieved with this method.

\subsection{Simulated training data\label{sec:simulated-training-data}}
To create a training and evaluation dataset for tritium detection, we simulated both tritium signal events and background events. For tritium, we generated 5~million electrons sampled from the tritium spectrum, emitted isotropically within a uniform 3\,\si{\micro\meter} 
 layer of ice on the CCD surface. 

The dominant background in the energy regime of tritium stems from relatively higher energy gamma rays that scatter with extremely gentle Compton interactions, depositing less than 20\,keV in the sensor. Though knowledge of the rate and energy spectrum of the incoming background gamma rays would be important for a detailed prediction of the observed rate in an experiment, we did not seek to model these aspects. Instead, we focused simply on generating a large sample of background events for algorithm training, and instead take the absolute background rates from experimental measurements. Towards this end, we simulated 20~million 1\,MeV gammas, and filtered the small fraction of these that yielded tritium-like energy depositions (greater than 0 and less than 20\,keV.) The energy spectrum of the resulting sample depends only weakly on the initial gamma ray energy, and we take these as representative training examples for gammas spanning a wide range of initial energies.

The process to transform simulated energy depositions into pixel-level training data follows multiple steps. First, we obtain the energy deposition (beyond the deadlayer) from Geant4. Then, we model the drift and diffusion of the liberated charges using our Python implementation described earlier. These charges are binned into 10$\times$10\,\si{\micro\meter}$^2$ pixels, matching the physical dimensions of our CCD. For the deep learning models, the pixel values are normalized so that all input values fall between 0 and 1.

We only keep clusters with at least one pixel above 4$\sigma$ of the measured noise (51\,eV). Clusters with an integrated energy above 20\,keV are discarded as they can be disregarded as not being tritium signals with a simple energy cut. After this selection process, we obtained approximately 1~million distinct clusters from each category (tritium and background) that were divided into a 70/20/10 split between training, validation, and testing sets.

\section{Analysis Methods}

Due to the short range of tritium beta particles, clusters produced from tritium and background show large differences in diffusion profiles. Tritium beta ray interactions are limited to the exposed surface of the CCD, and the charge produced must drift across the entire sensor bulk, diffusing maximally during the journey to the pixelated readout plane. Background events from gamma rays, on the other hand, interact throughout the entire CCD volume, yielding a wide range of drift lengths and resulting cluster sizes. This shape difference is the basis for much of the discrimination power between the tritium signal and background.

 In this section, we present several approaches to classify clusters as tritium or background that exploit distinguishing features like the particle path, energy deposition, and spatial charge spread. We show a classical technique that selects clusters based on their width and energy, decision trees that optimize the selection further, and deep learning architectures that learn to classify by looking at entire clusters. Each method aims to maximize the separation between tritium events and background events, ultimately enhancing the sensitivity of the system as a deployable tritium detector.

\subsection{Minimum Detectable Activity}
\label{sec:analysis}
To quantitatively evaluate and compare the different tritium classifier approaches, we employ the concept of Minimum Detectable Activity (MDA). The MDA is related to the Minimum Detectable Signal (MDS), defined by the Currie equation for a single bin counting experiment~\cite{Currie1968}. The MDS represents the smallest signal yield that can be detected with 5\% false negative rate, for a detection threshold with 5\% false positive rate. It is given by

\begin{equation}
    \text{MDS} = 4.653 \cdot \sigma_{\text{bkg}} + 2.706.
    \label{eq:mds}
\end{equation}

The MDS comes out to $\sim$30-170 tritium tracks that deposit energy in the detector per hour for the detector we used and background rate we measured in this study, depending on the classifier. To convert the MDS to the lowest activity of tritium that can be detected, it is scaled by the measurement time, $t$, and the efficiency for tritium detection, factorized in two parts: the classifier True Positive Rate (TPR) and the tritium deposition factor, $f_{t}$. This yields the MDA,  
\begin{equation}
    \text{MDA} = \frac{\text{MDS}}{f_{t} \cdot t \cdot \text{TPR}}.
     \label{eq:mda}
\end{equation}

The tritium deposition factor, $f_{t}$, represents the fraction of tritium beta rays that will reach the sensitive volume of the CCD from an ice layer with thickness greater than the saturating thickness of 3\,microns. We express this factor in units convenient for interpretation of small quantities from aqueous samples. Based on simulations described in Section~\ref{sec:ice-layer}, this factor is found to be 1.57~counts per hour, per mBq/µl of activity in the aqueous sample, for a sensor with active area of 15.84\,cm$^2$. The TPR is simply the fraction of the events reaching the sensitive volume that are classified as tritium. The MDA is then ultimately expressed as a volumetric specific activity, in units of mBq/µl, with assumptions of a 24\,hour counting time, a sensor of 15.84\,cm$^2$ active area, and an ice layer at least 3\,microns thick. 
The detection method exhibits linear response from the MDA up to approximately 15--40\,Bq/µl, above which cluster pileup (tritium-tritium, or tritium-background overlaps) in the CCD would begin to affect linearity. Overlapping tracks are not expected to be recoverable. The impact of pileup on the sensitivity can be incorporated by updating the effective active area to exclude areas covered by previous particle tracks in the same exposure.


The Poisson standard deviation of background counts appearing in Eq.~\ref{eq:mds}, $\sigma_{\text{bkg}}$, is given by:

\begin{equation}
    \sigma_{\text{bkg}} = \sqrt{R_{\text{bkg}} \cdot t \cdot \text{FPR}}.
\end{equation}

Here, $R_{\text{bkg}}$ is the rate of background events  with energy deposition compatible with tritium beta rays, (taken as 240/h as observed in our experimental data), and FPR is the false positive rate of the classifier.

For each classifier, we calculate the true positive rate (TPR) and false positive rate (FPR) across various discrimination thresholds. These rates directly impact the detection sensitivity: a higher TPR improves signal efficiency, while a lower FPR reduces background contamination. The optimal threshold is selected that results in the lowest MDA, balancing the trade-off between signal efficiency and background rejection.

\subsection{Classical Approach}

\begin{figure}[ht]
    \centering
    \includegraphics[width=0.95\linewidth]{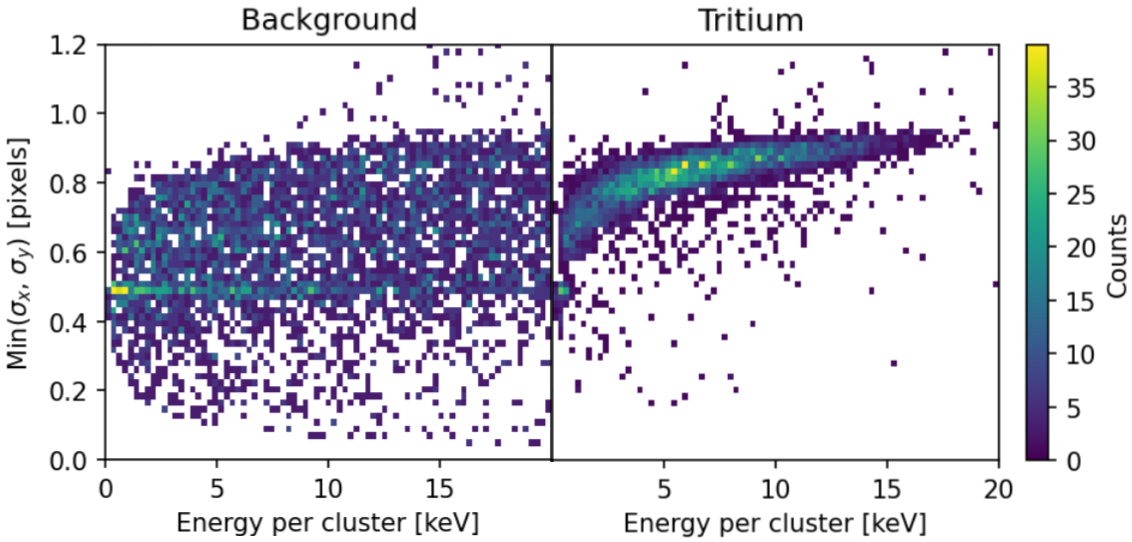}
    \caption{Measured background and tritium, plotted by the minimum width and energy per cluster.}
    \label{fig:classic}
\end{figure}

Inspired by the CCD analysis for the DAMIC dark matter experiment~\cite{aguilar-arevalo_search_2016, aguilar-arevalo_results_2020}, we implement a classical analysis based on simple variables formed from each cluster. This approach uses three variables: the cluster energy sum, and the widths along the X and Y dimensions, taken as the standard deviation of the projections of all pixels above threshold to the X and Y axes. Fig.~\ref{fig:classic} shows the energy (integral of the pixels above 4$\sigma$) versus spread (minimum of the widths in X and Y) for both the background and tritium experimental datasets. 
For such low energy interactions, the path of the tritium beta ray or scattered electron is limited to a length less than one pixel. The observed cluster shapes, then, are dominated by the transverse diffusion that occurs as the charges drift to the pixelated readout.
As tritium interactions occur only at the exposed surface, the resulting charge has to drift across the entire sensor volume, experiencing large diffusion due to the long drift. These events occupy a narrow band in Fig.~\ref{fig:classic} with large width. The background events, in contrast, interact essentially uniformly throughout the bulk, resulting in a broad distribution of widths.

To form the classical discriminator, bins from the histogram in Fig.~\ref{fig:classic} can be selected based on the concentration of tritium events, or alternatively, the bin-by-bin tritium to background ratio. Events are given a discriminator score based on their location within this space. Bins with higher tritium content are given a higher score, and the scores are mapped so that tritium events uniformly populate the range from 0 to 1. A requirement of discriminator value greater than fraction $f$ then selects events corresponding to the smallest region from Fig.~\ref{fig:classic} that includes $1-f$ of the tritium sample. 




\subsection{Boosted Decision Trees (BDTs)}

To optimize use of the cluster analysis variables introduced for the classical analysis, we implemented a Boosted Decision Tree (BDT)~\cite{coadou_boosted_2022}.
BDTs are a machine learning approach that combines multiple simple decision trees into a strong ensemble classifier. Decision trees themselves are intuitive models that make predictions through a series of if-else rules (e.g., "if energy $>$ 5\,keV and width $<$ 1.2\,pixels, then classify as background"). Unlike deep neural networks, BDTs operate on a set of engineered features rather than raw pixel data, potentially offering computational advantages and improved interpretability.

For our BDT implementation, we utilized the GradientBoostingClassifier from scikit-learn\cite{Buitinck2013} with 100~estimators and a max depth of 3. The model was trained on the same three key features extracted from each cluster as the classical method: the cluster energy sum, and the standard deviation of pixel values along X and Y. These features capture the essential characteristics that differentiate tritium events from background: energy deposition and spatial spread. The BDT learns optimal decision boundaries in this three-dimensional feature space, more effectively separating the two classes than in the histogram-based classical approach.
The BDT training was performed with a Friedman MSE loss function, learning rate of 0.05, and stopping triggered after 20 iterations with change in loss less than 0.0001.
Many variations in the input features and hyperparameters were considered. The most significant aspect are the choice of input features. Discrimination power was significant reduced by excluding the cluster energy sum or limiting the width information to a single dimension as in the classical method (either $
\sigma_x, \sigma_y, \text{min}(\sigma_x,\sigma_y), \text{ or } \text{max}(\sigma_x,\sigma_y)$). Additional features including total number of pixels, mean energy per pixel, total widths in x and y were incorporated, but did not have any significant impact on the performance, likely because they are somewhat degenerate with the other input features. Similarly, parameter scans in the number of estimators and the max tree depth were not found to only weakly affect the performance in the neighborhood of the selected values.

The BDT outputs a probability score between 0 and 1, representing the likelihood that a given cluster originated from tritium decay. By adjusting the threshold on this score, we can tune the classifier to achieve the desired balance between true positive rate and false positive rate, optimizing for minimum detectable activity.

\subsection{Deep Learning Approaches}
Here we attempt to use deep neural networks to improve signal-to-background separation. These methods take in all or a subset of the pixel values in a cluster, using more of the available information than the non-deep learning methods described above.

\begin{figure*}[ht]
    \centering
    \includegraphics[width=0.95\linewidth]{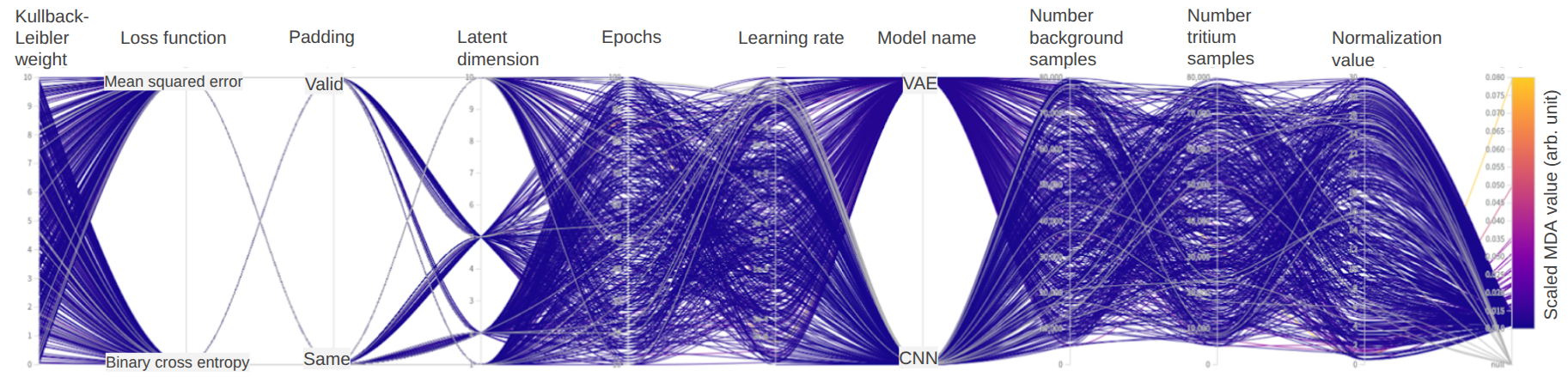}
    \caption{Parallel coordinates plot showing a hyperparameter sweep for two of the deep learning models. Each line represents one trained model configuration, colored by the final MDA achieved. Grey lines are models that failed to result in a valid MDA, for instance by labeling all events as background.}
    \label{fig:sweep}
\end{figure*}

To find the hyperparameters for each deep learning model that resulted in the best performance (lowest MDA), we performed random hyperparameter sweeps using Weights and Biases~\cite{wandb}. We ran multiple sweeps, each running for approximately 24~hours, training and evaluating thousands of different configurations. The hyperparameters included loss functions, learning rates, model sizes, and number oftraining epochs to identify optimal configurations. The objective for all sweeps was to minimize the MDA on the validation dataset. Fig.~\ref{fig:sweep} shows a parallel coordinates plot from one such sweep, illustrating how different hyperparameter combinations affect the final MDA. The best-performing hyperparameters identified through these sweeps are reported in the tables accompanying each model architecture below.

\subsubsection{Convolutional Neural Networks (CNNs)}
\begin{figure}[ht]
    \centering
    \includegraphics[width=0.85\linewidth]{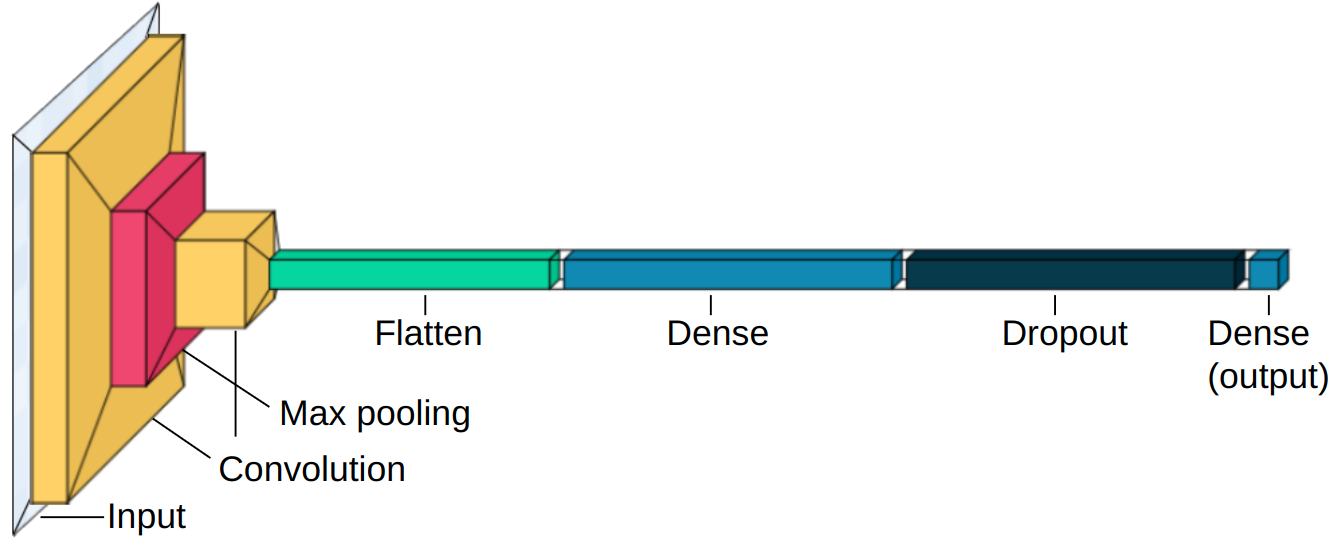}
    \caption{Schematics of the CNN architecture built and trained for this study.}
    \label{fig:cnn-model}
\end{figure}

CNNs~\cite{lecun_deep_2015} are a type of deep neural network that are particularly effective for image classification by learning spatial hierarchies of features from data. A CNN architecture, shown in Fig.~\ref{fig:cnn-model}, was trained on the labeled dataset containing both tritium and background events. The CNN takes in the 10$\times$10 pixel cluster image and outputs a single number between 0 and 1 indicating the confidence that the cluster comes from tritium. The window size of 10$\times$10, approximately a factor of 5 larger than full width of the largest tritium clusters, was selected to comfortably contain all tritium and tritium-like clusters. Two convolutional layers are used to extract features from the image, which are then flattened into an X long vector. This vector is then passed through a dense (fully connected) layer to output the single prediction number. A dropout layer is used to prevent overfitting by randomly dropping out 20\% of the neurons during training. The total number of trainable parameters in the model is 9697. The model was trained for 50~epochs with a learning rate of 0.001. A training typically takes $\sim$2 minutes to complete on an RTX 4090 GPU. The hyperparameters that achieved the lowest MDA through the sweep are summarized in Table~\ref{tab:cnn_hyperparams}.

\begin{table}[ht]
\centering
\caption{Lowest MDA CNN Hyperparameters}
\label{tab:cnn_hyperparams}
\begin{tabular}{lc}
\toprule
\textbf{Hyperparameter} & \textbf{Value} \\
\midrule
Convolutional Layer 1 & 8 filters, 3$\times$3 kernel \\
Convolutional Layer 2 & 16 filters, 3$\times$3 kernel \\
Dense Layer Size & 128 \\
Dropout Rate & 0.2 \\
Learning Rate & 0.001 \\
Batch Size & 1024 \\
Epochs & 50 \\
Activation Function & ReLU \\
Optimizer & Adam \\
Loss Function & Binary Focal Crossentropy ($\gamma=2$) \\
Padding & "same" \\
\bottomrule
\end{tabular}
\end{table}

\subsubsection{Particle Flow Networks (PFNs)}
Particle Flow Networks (PFNs) is a deep learning approach originally developed for high-energy physics applications~\cite{Komiske2019}. Unlike CNNs, which process the entire image grid, PFNs operate on a set representation of the data, treating each cluster as a collection of individual pixels with specific properties.

The PFN architecture is based on the Deep Sets framework, which provides a principled approach for processing unordered sets of objects.

Our implementation processes each cluster by first identifying a list of pixels above a 4$\sigma$ noise threshold, then representing each pixel as a feature vector containing the pixel energy deposition, and the x and y coordinates, defined relative to the brightest pixel of the cluster. In contrast to the CNN, the PFN only intakes the set of pixels above this noise threshold, rather than a fixed size grid of pixel energy values. This difference may sacrifice some discrimination power found in the near-threshold pixels at the periphery of the cluster, but also makes the classifier significantly less sensitive to drifts in detector noise relative to the training data.

The PFN architecture consists of two neural networks: a per-pixel network $\Phi$ that processes each pixel independently, and a global network $F$ that aggregates these features through a permutation-invariant operation (typically summation, as in our case) to produce the final classification output. Our implementation uses three dense layers for both networks with sizes (128, 128, 128) and (100, 100, 2) respectively, resulting in approximately 50,000 trainable parameters. Sweeps of the size of the $\Phi$ and F layers were performed, and were not found to strongly affect the performance in the neighborhood of the selected values.
The hyperparameters that achieved the lowest MDA through the sweep are summarized in Table~\ref{tab:pfn_hyperparams}.

\begin{table}[ht]
\centering
\caption{Lowest MDA PFN Hyperparameters}
\label{tab:pfn_hyperparams}
\begin{tabular}{lc}
\toprule
\textbf{Hyperparameter} & \textbf{Value} \\
\midrule
$\Phi$ Network Layers & (128, 128, 128) \\
$F$ Network Layers & (100, 100, 2) \\
Learning Rate & 0.0005 \\
Batch Size & 128 \\
Epochs & 75 \\
Activation Function & ReLU \\
Optimizer & Adam \\
Aggregation Function & Sum \\
Input Features & 3 (E, x, y) \\
\bottomrule
\end{tabular}
\end{table}

\subsubsection{Autoencoders}
Previous work has shown that unsupervised models like autoencoders can be trained exclusively on a single class, and still be effective for binary classification\cite{Farina2020,Dillon2022}. In this modality, the models are trained exclusively on background events and flag unfamiliar events as interesting anomalies. This strategy is particularly appealing when searching for generic or unspecified new signals, as it requires only minimal assumptions on the signal properties.

For the task of tritium detection in CCDs though, the anomaly detection is useful in the opposite perspective. In this case, generation of large signal samples for training is relatively easy, using a high activity source. The background, on the other hand, is intrinsically rare, of nebulous origin, and potentially variable. It is difficult to artificially enhance the background rate without modifying the background properties. It is therefore natural to train a model to gain familiarity with tritium signals, and classify non-tritium-like anomalies as background.

\begin{figure}[ht]
    \centering
    \includegraphics[width=0.70\linewidth]{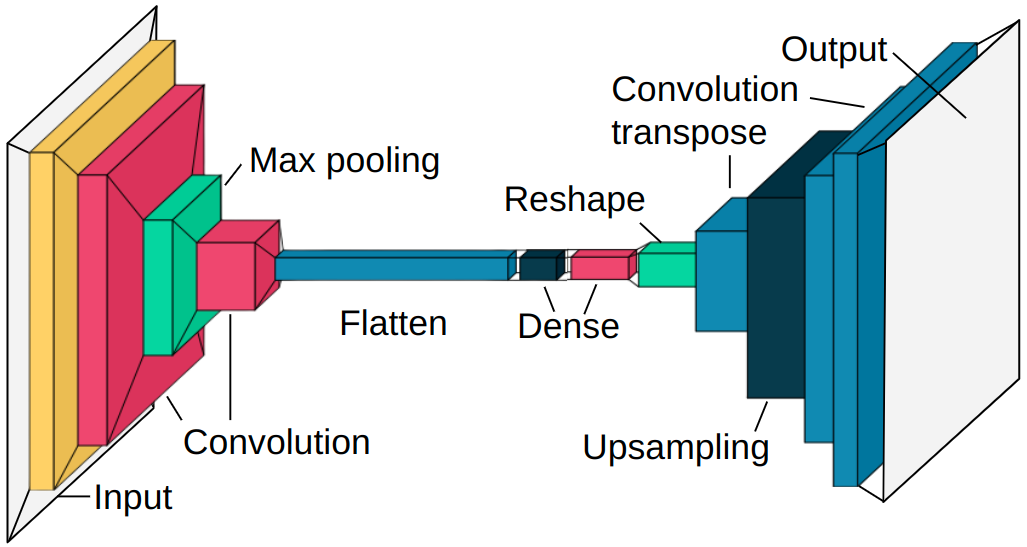}
    \caption{Schematics of the autoencoder architecture built and trained for this study.}
    \label{fig:autoencoder}
\end{figure}

A variational autoencoder (VAE) architecture was used for this purpose.  The VAE model, shown in Fig.~\ref{fig:autoencoder}, first encodes input images into a compressed latent representation with a defined probability distribution (characterized by mean and variance parameters) and then reconstructs them back into output images. When trained exclusively on tritium samples, the model becomes highly specialized at reproducing the distinctive features of tritium tracks.

\begin{figure}
    \centering
    \includegraphics[width=0.45\linewidth]{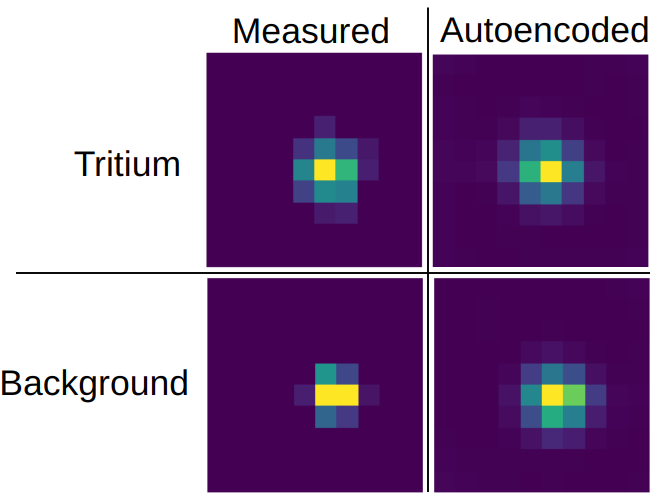}
    \caption{Measured particle tracks from tritium and a background event. The right tracks are generated by an autoencoder that is trained only on tritium tracks. The autoencoder is behaving as designed and is better at recreating the tritium track than the track from a background event. In other words, the mean squared error between the measured and autoencoded tritium images are lower than for the background images.}
    \label{fig:grid}
\end{figure}

This specialization creates a natural discriminator: when presented with background events, the model produces significantly higher reconstruction errors (quantified by the mean squared error) compared to tritium events, as illustrated in Fig.~\ref{fig:grid}. We leverage this difference in reconstruction quality to classify events, using the mean squared error between input and output images as our classification metric.

What distinguishes a VAE from a standard autoencoder is the addition of Kullback-Leibler divergence as a regularization term during training. This term constrains the latent space distribution to approximate a standard normal distribution, which helps prevent overfitting and improves the model's ability to generalize. The hyperparameters that achieved the lowest MDA through the sweep are summarized in Table~\ref{tab:vae_hyperparams}.

\begin{table}[ht]
\centering
\caption{Lowest MDA VAE Hyperparameters}
\label{tab:vae_hyperparams}
\begin{tabular}{lc}
\toprule
\textbf{Hyperparameter} & \textbf{Value} \\
\midrule
Encoder Conv Layers & 8, 16 filters (3$\times$3 kernel) \\
Latent Dimension & 5 \\
Decoder Conv Layers & 16, 8 filters (3$\times$3 kernel) \\
Learning Rate & 0.001 \\
Batch Size & 256 \\
Epochs & 60 \\
Kullback-Leibler Weight & 0.2 \\
Loss Function & Mean Squared Error \\
Optimizer & Adam \\
Padding & "same" \\
\bottomrule
\end{tabular}
\end{table}

\section{Results}

\begin{figure}
    \centering
    \includegraphics[width=0.9\linewidth]{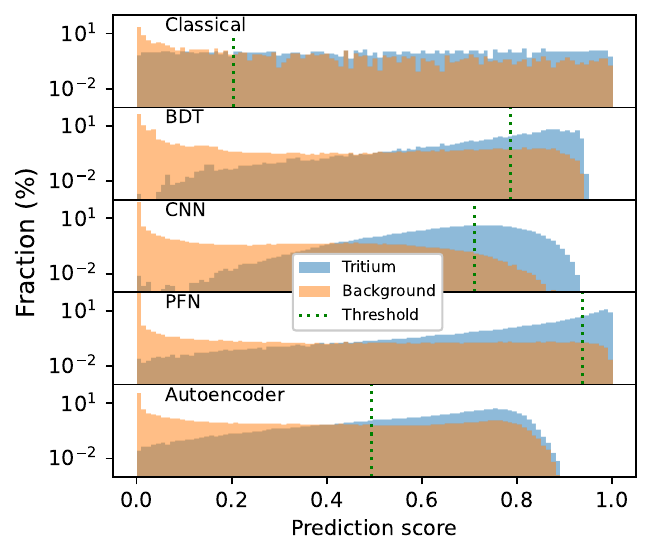}
    \caption{Comparison of discriminator scores for the different methods explored in this paper: Classical approach, BDT, CNN, PFN, and Autoencoder. For each method, the distributions show the scores for tritium events (blue) and background events (orange) The threshold that results in the lowest MDA is marked with a vertical dashed line.}
    \label{fig:discriminator_scores}
\end{figure}

The results from each approach are summarized in Fig.~\ref{fig:discriminator_scores}, which shows the discriminator scores for each method when applied to tritium and background events. The thresholds that result in the lowest MDA for each method are marked with a vertical line.

The classical approach (Fig.~\ref{fig:discriminator_scores}a) shows reasonable separation between tritium and background events, with most background events scoring below 0.2 while tritium scores are uniformly distributed by construction. This method provides a solid baseline but leaves room for improvement.

Results from the BDT  (Fig.~\ref{fig:discriminator_scores}d) achieve good separation between tritium and background events, with an area under the receiver operating characteristic (ROC) curve of approximately 0.94. At an operating threshold of 0.877, the BDT maintains a high true positive rate while reducing false positives compared to the classical approach.

Our CNN approach (Fig.~\ref{fig:discriminator_scores}b) demonstrates superior discrimination capability compared to the classical method. The CNN clearly assigns higher scores to tritium events and lower scores to background events, indicating it has effectively learned the distinctive features that separate these two classes. The PFN achieves similar discrimination power, but with scores that are bunched closer to 1 for both tritium and background events, and a correspondingly higher optimum threshold.

The autoencoder (Fig.~\ref{fig:discriminator_scores}c) achieves discrimination by using reconstruction error as a metric—events that deviate from the learned tritium patterns produce higher errors. Its performance is notably weaker than the supervised methods (CNN, PFN, and BDT). This performance gap stems from a fundamental difference: the autoencoder learns only from tritium samples, while supervised methods leverage labeled examples from both classes to directly learn the decision boundary. This results in a substantially higher false positive rate for the autoencoder. Despite this limitation, the background-agnostic training approach offers potential advantages when background characteristics are poorly defined or variable.

\begin{figure}
    \centering
    \includegraphics[width=0.85\linewidth]{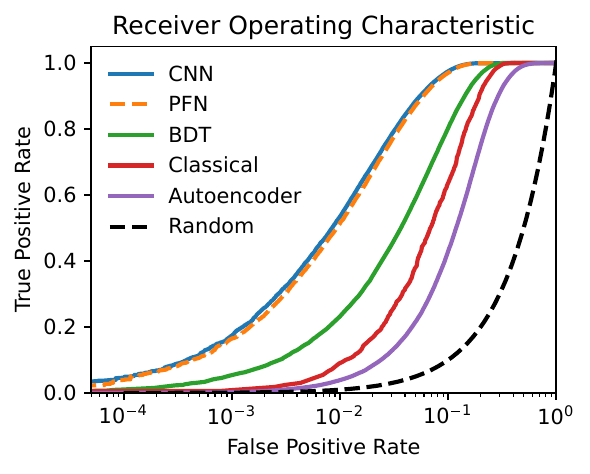}
    \caption{ROC curves for the different methods. A random classifier is shown as a dashed line for reference. Better performance is indicated by a higher area under the curves with the CNN and PFN showing the best performance.}
    \label{fig:roc}
\end{figure}

Fig.~\ref{fig:roc} presents ROC curves comparing all tested methods. The CNN and PFN provide the best overall performance, with almost identical ROC curves. The non-deep BDT and classical classifiers come next, and the autoencoder shows the worst classification performance of the methods tested. Relative to the BDT, the deep methods provide a factor of 5-6~stronger background rejection at the same tritium efficiency.

Table~\ref{tab:method_comparison} compares the performance of the various methods optimized to minimize the MDA in a 24 hour counting experiment. The true and false positive rates (TPR and FPR) are presented at the optimum working point, as well as the corresponding number of false positives (FP) for the incident background rate of 240 events per hour based on the experimental exposures. The resulting minimum detectable signal yield (MDS) and activity (MDA) are reported, based on Eqs.~\ref{eq:mds} and ~\ref{eq:mda}.

\begin{table*}[ht]
\centering
\caption{Comparison of Classification Methods}
\label{tab:method_comparison}
\begin{tabularx}{0.65\textwidth}{lXcccccccc}
\toprule
\textbf{Method} & \textbf{MDA} & \textbf{Training} & \textbf{Params} & \textbf{TPR} & \textbf{FPR} & \textbf{FP} & \textbf{MDS} & \textbf{AUC} \\
                & (mBq/µl)     & \textbf{Time}     & ($10^3$)            & (\%)         & (\%)   & (per 24\,h)& (per 24\,h) &      \\
\midrule
Classical      & 4.6 & 1\,s   & 42.3 & 80.8 & 14.8 & 852 & 139 & 0.9108 \\
BDT            & 3.7 & 4\,min & 3    & 65.3 & 6.2  & 357 & 91 & 0.9433 \\
CNN            & 1.8 & 2\,min & 10   & 43.7 & 0.6 & 35 & 30 & 0.9804 \\
PFN            & 1.9 & 6\,min & 56.5 & 48.4 & 0.9 & 52 & 36 & 0.9789 \\
Autoencoder   & 5.6 & 5\,min & 12   & 82.4 & 23.5 & 1353 & 174 & 0.8583 \\
\bottomrule
\end{tabularx}
\end{table*}

\begin{figure}[ht]
    \centering
    \includegraphics[width=0.85\linewidth]{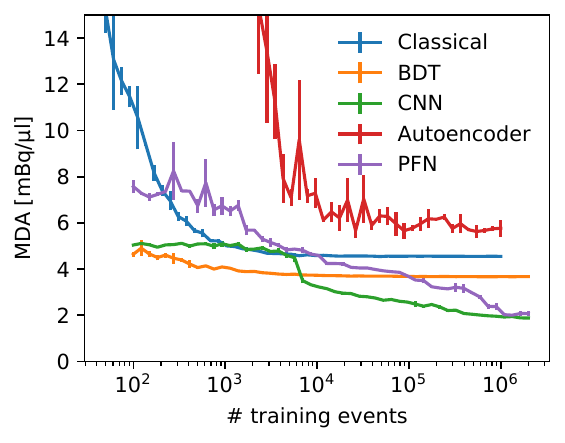}
    \caption{Minimum detectable concentration of water sample as a function of training events for the different classifiers. The non-deep classical and BDT methods don't benefit from adding more data after around 1000~images whereas the deep methods continue to improve with up to millions of training images.}
    \label{fig:training_mda}
\end{figure}

In Fig.~\ref{fig:training_mda}, we study the impact on training statistics on the classifier performance. Each classifier is trained with subsets of the simulated data, ranging from only 100~training images to the full 2~million training images of both background and tritium. The autoencoder, following its unsupervised approach, is trained only on tritium samples with the same range of training set sizes. In all cases, the evaluation uses the full validation dataset. Error bars show the standard deviation on the achieved MDA between 4~trainings with randomized subsets of the training data. All classifiers get better with more training data but we can observe how the deep models (CNN, autoencoder, and PFN) are able to utilize the information in high numbers of training examples, whereas the non-deep models (classical and BDT) plateau after around 10,000~training examples. The PFN lags behind the CNN but reach the same $\sim$2\,mBq/µl MDA on the full dataset. The lag could be due to the PFN having more trainable parameters than the CNN. An interesting observation is that the BDT performs the best up until about 5,000~training examples, after which the CNN has learned to classify background and tritium well enough to achieve the lowest MDA. The autoencoder plateaus at a higher MDA, reflecting its single-class training limitation. The parameter sizes reported in Table~\ref{tab:method_comparison} represent the configurations that resulted in the lowest MDA through the hyperparameter sweeps. Training times are provided for reference and were run on a single RTX 4090 GPU with the number of epochs determined through the hyperparameter sweeps.
Based on this analysis, we conclude training samples should contain at least 10,000 events from each category, and ideally 100,000, to provide an adequate number of examples for the deep learning methods. In simulation, tritium and background samples of this size are easily generated. Large experimental tritium samples are also easily generated with high activity tritium sources. On the other hand, collection of large experimental background samples is less convenient, but still feasible. At the observed background rate of roughly 200 low energy tritium candidates per hour, 10,000 events are obtained in roughly 50 hours of background exposure. Future work in a high background environment with additional gamma sources is anticipated, to accelerate collection of even larger samples. 

\begin{figure}
    \centering
    \includegraphics[width=0.9\linewidth]{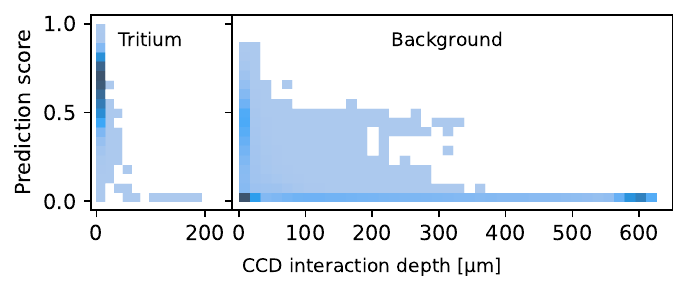}
    \caption{Heatmap of predictions from the CNN for tritium and background events. Darker blue indicates more events from the validation dataset. The x-axis shows the depth of particle interaction in the CCD and the y-axis shows the discriminator score (0 for background and 1 for tritium).}
    \label{fig:score_depth}
\end{figure}

In Fig.~\ref{fig:score_depth} we examine the best performing classifier, the CNN, more closely. By looking at the events that result in outlier predictions, we identified the depth of interaction of the particle in the CCD as the most important feature of the particle tracks for the CNN prediction. This is because the depth of interaction translates to spread of the charges as explained in Section~\ref{sec:diffusion}. The majority of the tritium events only deposit energy in the first \si{\micro\meter} of the CCD, and are given scores closer to 1. The few events with deeper interactions come from photons that are created from the tritium electron through Bremsstrahlung or characteristic x-rays, and result in clusters and subsequent prediction scores that look like background events. The background events are mainly from Compton scatterings where the incoming gamma ray photon freed an electron with energy in the tritium range. For Compton electrons that were freed deep within the CCD, the CNN has no problem correctly classifying the clusters as background, as seen in the distribution of prediction scores around 0 for the majority of background interactions deeper than $\sim$10\,\si{\micro\meter} in the CCD. The background events that interacted closer to the surface of the CCD are more difficult to classify and make the bulk of the false positive predictions. Ultimately, deep learning allows distinguishing events that are closer and closer to the surface, but still struggles for background interactions with exceedingly similar energy and depth as tritium beta rays.

As for the deployment of the computational models, the relatively small data amount of data generated (few GB per day) and use of pre-trained models will enable the daily quantification analysis to be executed comfortably and promptly even with modest in situ computing resources. 

\begin{figure}
    \centering
    \includegraphics[width=0.85\linewidth]{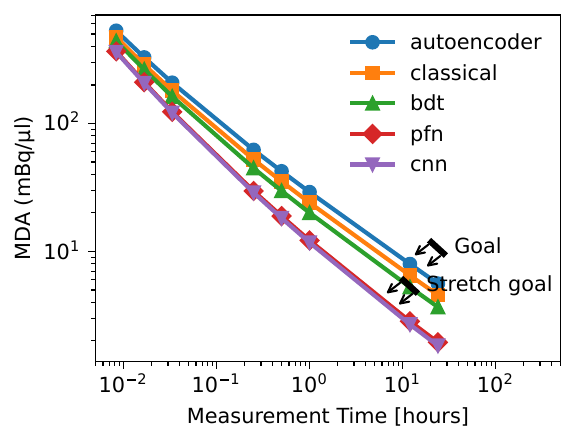}
    \caption{The MDA that can be achieved using the 5~different classifiers as a function of counting time. The GRAIL-defined goal and stretch goals for 24 and 12~hour measurements are indicated.}
    \label{fig:mda_vs_time}
\end{figure}

Finally, we consider the performance of the CCD analysis in the context of a gaseous environmental monitoring system, as envisioned in the GRAIL tritium program~\cite{grail}. In GRAIL, sensitivity targets were set for tritium concentration per cubic meter of air (SCM): 1\,pCi for a 36 hour measurement (Phase I), 0.1\,pCi in 24 hours (Phase II), and 0.05\,pCi in 12 hours (Stretch Goal). Supposing the presence of tritiated hydrogen gas (HT) with that specific activity, and regular hydrogen gas (H2) at typical atmospheric concentration (84\,g/ SCM), isolation and subsequent oxidation of both species together would yield water samples with activity of about 2.9\,pCi/µL, or 107\,mBq/µL (Phase I) and 10.7\,mBq/µL (Phase II), and 5.4\,mBq/µL (Stretch). The Phase II and Stretch goals specific activities are marked in Fig.~\ref{fig:mda_vs_time}. The deep learning CNN and PFN models are able to achieve even the stretch goals, while the Phase I and Phase II goals are attained comfortably even by relatively less performant discrimination methods. We conclude that a CCD-based analysis would be a strong contender for tritium quantification in a deployable gaseous detection system. 

\section{Conclusions}

We conclude that CCD-based detection offers a compelling alternative to traditional tritium detection methods due to the sensor's high resolution and efficiency for ultra-low energy beta rays.
We developed a realistic simulation of the tritium measurement process in CCD, including the effects of beta ray attenuation within the ice sample and within the inactive layers of the CCD, as well as the signal formation and diffusion of charge within the CCD. 
A variety of analysis methods were developed, including, two shallow classifiers (classical and BDT), and deep learning techniques (CNN, PFN, and autoencoder). We found that our CNN and PFN deep learning methods yielded the best performance in classifying tritium events, when given enough training data. The deep models could detect tritium in a water sample frozen onto a CCD detector at down to 1.8\,mBq/µl. That is a significant improvement in sensitivity over the the best non-deep classifier, the BDT, at 3.7\,mBq/µl. These sensitivities exceed the requirements set by the GRAIL tritium detection program.

Collecting the millions of background training examples needed to achieve that level of performance in the CNN or PFN classifiers could be a time-consuming process. In these cases background-agnostic classifiers like the autoencoder or the classical method could be valuable.

Moving forward, further refinement of the simulation models like configurations of detector thickness, size, and operation voltages, and extended benchmarking of additional machine learning approaches will be pursued. These studies will be critical for developing portable and robust tritium monitoring systems. Further work is also under way to realize these techniques in experimental measurements with high statistics.

Supplemental rejection of the remaining difficult background events could be achieved with auxiliary gamma detectors that collect the scattered gamma rays associated with the low energy Compton scatters near the CCD surface. Synchronizing this kind of veto system, though, would likely only be feasible with a very high frame-rate CCD or an active readout CMOS pixel sensor that provides modest timestamping resolution for each interaction.

\bibliographystyle{IEEEtran}
\bibliography{export,odd_ones}

@article{hou_critical_2008,
	title = {Critical comparison of radiometric and mass spectrometric methods for the determination of radionuclides in environmental, biological and nuclear waste samples},
	url = {https://linkinghub.elsevier.com/retrieve/pii/S0003267007019964},
	doi = {10.1016/j.aca.2007.12.012},
	urldate = {2023-11-16},
	journal = {Analytica Chimica Acta},
	author = {Hou, Xiaolin and Roos, Per},
	year = {2008},

}

@article{aguilar-arevalo_results_2020,
	title = {Results on {Low}-{Mass} {Weakly} {Interacting} {Massive} {Particles} from an 11 kg d {Target} {Exposure} of {DAMIC} at {SNOLAB}},
	url = {https://link.aps.org/doi/10.1103/PhysRevLett.125.241803},
	doi = {10.1103/PhysRevLett.125.241803},
        journal = {Phys. Rev. Lett.},
        author = {Aguilar-Arevalo, A. and Amidei, D. and Bertou, X. and Butner, M. and Cancelo, G. and Castañeda Vázquez, A. and Cervantes Vergara, B. A. and Chavarria, A. E. and Chavez, C. R. and de Mello Neto, J. R. T. and D’Olivo, J. C. and Estrada, J. and Fernandez Moroni, G. and Gaïor, R. and Guardincerri, Y. and Hernández Torres, K. P. and Izraelevitch, F. and Kavner, A. and Kilminster, B. and Lawson, I. and Letessier-Selvon, A. and Liao, J. and Mello, V. B. B. and Molina, J. and Peña, J. R. and Privitera, P. and Ramanathan, K. and Sarkis, Y. and Schwarz, T. and Sengul, C. and Settimo, M. and Sofo Haro, M. and Thomas, R. and Tiffenberg, J. and Tiouchichine, E. and Torres Machado, D. and Trillaud, F. and You, X. and Zhou, J. and {DAMIC Collaboration}},
	year = {2020},
}

@article{aguilar-arevalo_search_2016,
	title = {Search for low-mass {WIMPs} in a 0.6 kg day exposure of the {DAMIC} experiment at {SNOLAB}},
	url = {https://link.aps.org/doi/10.1103/PhysRevD.94.082006},
	doi = {10.1103/PhysRevD.94.082006},
	journal = {Phys. Rev. D},
	author = {Aguilar-Arevalo, A. and Amidei, D. and Bertou, X. and Butner, M. and Cancelo, G. and Castañeda Vázquez, A. and Cervantes Vergara, B. A. and Chavarria, A. E. and Chavez, C. R. and de Mello Neto, J. R. T. and D’Olivo, J. C. and Estrada, J. and Fernandez Moroni, G. and Gaïor, R. and Guardincerri, Y. and Hernández Torres, K. P. and Izraelevitch, F. and Kavner, A. and Kilminster, B. and Lawson, I. and Letessier-Selvon, A. and Liao, J. and Mello, V. B. B. and Molina, J. and Peña, J. R. and Privitera, P. and Ramanathan, K. and Sarkis, Y. and Schwarz, T. and Sengul, C. and Settimo, M. and Sofo Haro, M. and Thomas, R. and Tiffenberg, J. and Tiouchichine, E. and Torres Machado, D. and Trillaud, F. and You, X. and Zhou, J. and {DAMIC Collaboration}},
	year = {2016},
}

@article{holland_fully_2023,
	title = {Fully depleted charge‐coupled device design and technology development},
	url = {https://onlinelibrary.wiley.com/doi/10.1002/asna.20230072},
	doi = {10.1002/asna.20230072},
	urldate = {2023-08-31},
	journal = {Astron Nachr},
	author = {Holland, Stephen E.},
	year = {2023},
}

@article{goldschmidt_veryfastccd_2023,
	title = {{VeryFastCCD}: a high frame rate soft {X}-ray detector},
	volume = {11},
	shorttitle = {{VeryFastCCD}},
	url = {https://www.frontiersin.org/articles/10.3389/fphy.2023.1285350/full},
	doi = {10.3389/fphy.2023.1285350},
	journal = {Front. Phys.},
	author = {Goldschmidt, Azriel and Grace, Carl and Joseph, John and Krieger, Amanda and Tindall, Craig and Denes, Peter},
	year = {2023},
}

@article{Buitinck2013,
   author = {Lars Buitinck and Gilles Louppe and Mathieu Blondel and Fabian Pedregosa and Andreas C Müller and Olivier Grisel and Vlad Niculae and Peter Prettenhofer and Alexandre Gramfort and Jaques Grobler and Robert Layton and Jake Vanderplas and Arnaud Joly and Brian Holt and Gaël Varoquaux},
   month = {9},
   title = {{API} design for machine learning software: experiences from the scikit-learn project},
   url = {https://arxiv.org/abs/1309.0238v1},
   year = {2013}
}

@article{Komiske2019,
   author = {Patrick T. Komiske and Eric M. Metodiev and Jesse Thaler},
   doi = {10.1007/JHEP01(2019)121},
   issn = {1029-8479},
   issue = {1},
   journal = {Journal of High Energy Physics 2019 2019:1},
   month = {1},
   pages = {1-46},
   publisher = {Springer},
   title = {Energy flow networks: deep sets for particle jets},
   volume = {2019},
   url = {https://link.springer.com/article/10.1007/JHEP01(2019)121},
   year = {2019}
}

@article{Currie1968,
   author = {Lloyd A. Currie},
   doi = {10.1021/AC60259A007/ASSET/AC60259A007.FP.PNG_V03},
   issn = {15206882},
   issue = {3},
   journal = {Analytical Chemistry},
   month = {3},
   pages = {586-593},
   publisher = {American Chemical Society},
   title = {Limits for Qualitative Detection and Quantitative Determination: Application to Radiochemistry},
   volume = {40},
   url = {https://pubs.acs.org/doi/abs/10.1021/ac60259a007},
   year = {1968}
}

@article{Dillon2022,
   author = {Barry M. Dillon and Radha Mastandrea and Benjamin Nachman},
   doi = {10.1103/PHYSREVD.106.056005/FIGURES/12/MEDIUM},
   issn = {24700029},
   issue = {5},
   journal = {Physical Review D},
   month = {9},
   pages = {056005},
   publisher = {American Physical Society},
   title = {Self-supervised anomaly detection for new physics},
   volume = {106},
   url = {https://journals.aps.org/prd/abstract/10.1103/PhysRevD.106.056005},
   year = {2022}
}

@article{Farina2020,
   author = {Marco Farina and Yotam Nakai and David Pardo and David Shih},
   doi = {10.1103/PhysRevD.101.075021},
   issn = {24700029},
   issue = {7},
   journal = {Physical Review D},
   month = {4},
   pages = {075021},
   publisher = {American Physical Society},
   title = {Searching for new physics with deep autoencoders},
   volume = {101},
   url = {https://journals.aps.org/prd/abstract/10.1103/PhysRevD.101.075021},
   year = {2020}
}

@misc{NIST,
   doi = {10.18434/T4NC7P},
   title = {{Stopping-Power and Range Tables for Electrons, Protons, and Helium Ions | NIST}},
   url = {https://www.nist.gov/pml/stopping-power-range-tables-electrons-protons-and-helium-ions}
}

@article{Janesick1987,
   author = {James R. Janesick and Tom Elliott and Stewart Collins and Morley M. Blouke and Jack Freeman},
   doi = {10.1117/12.7974139},
   issn = {0091-3286},
   issue = {8},
   journal = {https://doi.org/10.1117/12.7974139},
   month = {8},
   pages = {692-714},
   publisher = {SPIE},
   title = {Scientific Charge-Coupled Devices},
   volume = {26},
   year = {1987}
}

@article{britt2022,
   author = {C Britt and E Church and T Hossbach and B Loer and R Saldanha and N Sinha and K Woodruff},
   journal = {{arXiv}},
   month = {1},
   title = {Simulated Thick, Fully-Depleted CCD Exposures Analyzed with Deep Learning Techniques},
   url = {https://arxiv.org/abs/2201.08973v1},
   year = {2022}
}

@article{holland2003,
   author = {Stephen E. Holland and Donald E. Groom and Nick P. Palaio and Richard J. Stover and Mingzhi Wei},
   doi = {10.1109/TED.2002.806476},
   issn = {00189383},
   issue = {1},
   journal = {IEEE T-ED},
   month = {1},
   pages = {225-238},
   title = {Fully depleted, back-illuminated charge-coupled devices fabricated on high-resistivity silicon},
   volume = {50},
   year = {2003}
}

@article{cabello2010,
   author = {Jorge Cabello and Kevin Wells},
   doi = {10.1088/0031-9155/55/6/010},
   issn = {0031-9155},
   issue = {6},
   journal = {Physics in Medicine and Biology},
   month = {3},
   pages = {1677},
   pmid = {20197603},
   publisher = {IOP Publishing},
   title = {The spatial resolution of silicon-based electron detectors in beta-autoradiography},
   volume = {55},
   year = {2010}
}

@article{geant4_1,
   author = {S. Agostinelli and J. Allison and K. Amako and J. Apostolakis and H. Araujo and P. Arce and M. Asai and D. Axen and S. Banerjee and G. Barrand and F. Behner and L. Bellagamba and J. Boudreau and L. Broglia and A. Brunengo and H. Burkhardt and S. Chauvie and J. Chuma and R. Chytracek and G. Cooperman and G. Cosmo and P. Degtyarenko and A. Dell'Acqua and G. Depaola and D. Dietrich and R. Enami and A. Feliciello and C. Ferguson and H. Fesefeldt and G. Folger and F. Foppiano and A. Forti and S. Garelli and S. Giani and R. Giannitrapani and D. Gibin and J. J. Gomez Cadenas and I. Gonzalez and G. Gracia Abril and G. Greeniaus and W. Greiner and V. Grichine and A. Grossheim and S. Guatelli and P. Gumplinger and R. Hamatsu and K. Hashimoto and H. Hasui and A. Heikkinen and A. Howard and V. Ivanchenko and A. Johnson and F. W. Jones and J. Kallenbach and N. Kanaya and M. Kawabata and Y. Kawabata and M. Kawaguti and S. Kelner and P. Kent and A. Kimura and T. Kodama and R. Kokoulin and M. Kossov and H. Kurashige and E. Lamanna and T. Lampen and V. Lara and V. Lefebure and F. Lei and M. Liendl and W. Lockman and F. Longo and S. Magni and M. Maire and E. Medernach and K. Minamimoto and P. Mora de Freitas and Y. Morita and K. Murakami and M. Nagamatu and R. Nartallo and P. Nieminen and T. Nishimura and K. Ohtsubo and M. Okamura and S. O'Neale and Y. Oohata and K. Paech and J. Perl and A. Pfeiffer and M. G. Pia and F. Ranjard and A. Rybin and S. Sadilov and E. di Salvo and G. Santin and T. Sasaki and N. Savvas and Y. Sawada and S. Scherer and S. Sei and V. Sirotenko and D. Smith and N. Starkov and H. Stoecker and J. Sulkimo and M. Takahata and S. Tanaka and E. Tcherniaev and E. Safai Tehrani and M. Tropeano and P. Truscott and H. Uno and L. Urban and P. Urban and M. Verderi and A. Walkden and W. Wander and H. Weber and J. P. Wellisch and T. Wenaus and D. C. Williams and D. Wright and T. Yamada and H. Yoshida and D. Zschiesche},
   doi = {10.1016/S0168-9002(03)01368-8},
   issn = {01689002},
   issue = {3},
   journal = {Nucl. Instr. Meth. Phys. Res. A},
   month = {7},
   pages = {250-303},
   publisher = {North-Holland},
   title = {GEANT4 - A simulation toolkit},
   volume = {506},
   year = {2003}
}

@article{mao_measurement_2024,
	title = {Measurement {Techniques} for {Low}-{Concentration} {Tritium} {Radiation} in {Water}: {Review} and {Prospects}},
	url = {https://www.mdpi.com/1424-8220/24/17/5722},
	doi = {10.3390/s24175722},
	journal = {Sensors},
	author = {Mao, Junxiang and Chen, Ling and Xia, Wenming and Gong, Junjun and Chen, Junjun and Liang, Chengqiang},
	year = {2024},

}

@article{lin_electrolytic_2020,
	title = {Electrolytic enrichment method for tritium determination in the {Arctic} {Ocean} using liquid scintillation counter},
	url = {https://link.springer.com/10.1007/s13131-020-1647-4},
	doi = {10.1007/s13131-020-1647-4},

	journal = {Acta Oceanol. Sin.},
	author = {Lin, Feng and Yu, Tao and Yu, Wen and Ni, Jialin and Lin, Li},
	year = {2020},
}

@article{furuta_measurement_2014,
	title = {Measurement of tritium with high efficiency by using liquid scintillation counter with plastic scintillator},
	url = {https://linkinghub.elsevier.com/retrieve/pii/S0969804314001341},
	doi = {10.1016/j.apradiso.2014.04.001},
	journal = {Applied Radiation and Isotopes},
	author = {Furuta, Etsuko and Ohyama, Ryu-ichiro and Yokota, Shigeaki and Nakajo, Toshiya and Yamada, Yuka and Kawano, Takao and Uda, Tatsuhiko and Watanabe, Yasuo},
	year = {2014},
}

@article{plastino_tritium_2007,
	title = {Tritium in water electrolytic enrichment and liquid scintillation counting},
	url = {https://linkinghub.elsevier.com/retrieve/pii/S1350448706001193},
	doi = {10.1016/j.radmeas.2006.07.010},
	journal = {Radiation Measurements},
	author = {Plastino, Wolfango and Chereji, Iosif and Cuna, Stela and Kaihola, Lauri and De Felice, Pierino and Lupsa, Nicolae and Balas, Gabriela and Mirel, Valentin and Berdea, Petre and Baciu, Calin},
	year = {2007},
}

@article{street_ciency_nodate,
	title = {High Efficiency {Detection} of {Tritium} {Using} {Silicon} {Avalanche} {Photodiodes}},
	language = {en},
	author = {Shah, K.S. and Gothoskar, P. and Farrell, R. and Gordon, J.},
	journal = {IEEE Transactions on Nuclear Science},
	year = {1997},
}

@article{sanada_development_2024,
	title = {Development of a practical tritiated water monitor to supervise the discharge of treated water from {Fukushima} {Daiichi} {Nuclear} {Power} {Plant}},
	url = {https://linkinghub.elsevier.com/retrieve/pii/S0168900224001347},
	doi = {10.1016/j.nima.2024.169208},
	journal = {Nuclear Instruments and Methods in Physics Research Section A: Accelerators, Spectrometers, Detectors and Associated Equipment},
	author = {Sanada, Yukihisa and Oshikiri, Keisuke and Kanno, Marina and Abe, Tomohisa},
	year = {2024},
	pages = {169208},
}

@article{uda_detection_2010,
	title = {Detection efficiency of plastic scintillator for gaseous tritium sampling and measurement system},
	url = {https://linkinghub.elsevier.com/retrieve/pii/S0920379610001602},
	doi = {10.1016/j.fusengdes.2010.04.019},
	journal = {Fusion Engineering and Design},
	author = {Uda, Tatsuhiko and Kawano, Takao and Tanaka, Masahiro and Tomatsuri, Satoshi and Ito, Takeshi and Tatenuma, Katsuyoshi},
	year = {2010},
}

@textbook{horrocks_applications_2012,
	title = {Applications of {Liquid} {Scintillation} {Counting}},
	isbn = {9780123918600},
	url = {https://linkinghub.elsevier.com/retrieve/pii/B9780123918600000017},
	publisher = {Elsevier},
	author = {Horrocks, D.L.},
	year = {2012},
}

@article{hofstetter_field_1999-1,
	title = {Field deployable tritium analysis system for ground and surface water measurements},
	url = {https://link.springer.com/10.1007/BF02383549},
	doi = {10.1007/BF02383549},
	journal = {Journal of Radioanalytical and Nuclear Chemistry},
	author = {Hofstetter, K.J. and Cable, P.R. and Beals, D.M. and Noakes, J.E. and Spaulding, J.D. and Neary, M.P. and Peterson, R.N.},
	year = {1999},
}

@inproceedings{bowman_proportional_1981,
	title = {Proportional counting techniques for routine tritium analyses at environmental levels},
	booktitle = {IAEA international symposium on methods of low-level counting and spectrometry},
	author = {Bowman, W.W. and Hughes, M.B.},
	year = {1981},
	address = {Berlin, F.R. Germanty},
}

@article{stanga_improved_2006,
	title = {Improved method of measurement for tritiated water standardization by internal gas proportional counting},
	url = {https://linkinghub.elsevier.com/retrieve/pii/S0969804305002289},
	doi = {10.1016/j.apradiso.2005.06.009},
	urldate = {2025-10-01},
	journal = {Applied Radiation and Isotopes},
	author = {Stanga, D. and Cassette, P.},
	year = {2006},
}

@incollection{coadou_boosted_2022,
	title = {Boosted decision trees},
	url = {http://arxiv.org/abs/2206.09645},
	author = {Coadou, Yann},
	year = {2022},
	doi = {10.1142/9789811234033_0002},
}

@article{lecun_deep_2015,
	title = {Deep learning},
	url = {https://www.nature.com/articles/nature14539},
	doi = {10.1038/nature14539},
	journal = {Nature},
	author = {LeCun, Yann and Bengio, Yoshua and Hinton, Geoffrey},
	year = {2015},
}

@misc{grail,
    title={{GRAIL Program}},
howpublished={\url{https://www.iarpa.gov/research-programs/grail}},
    author={{Intelligence Advanced Research Projects Activity}}
}

@misc{wandb,
    title = {Experiment Tracking with Weights and Biases},
    year = {2020},
    note = {Software available from wandb.com},
    url={https://www.wandb.com/},
    author = {Biewald, Lukas}
}

\end{document}